\documentclass[runningheads]{llncs}
\pdfoutput=1
\usepackage{amsmath}
\usepackage{amsfonts}
\usepackage{siunitx}
\usepackage{mleftright}
\usepackage{import}
\usepackage{graphicx}
\usepackage{tikz}
\usetikzlibrary{positioning,arrows,matrix}

\usepackage{xcolor}

\begin{document}
\title{Topological descriptors of spatial coherence in a convective boundary layer}
\titlerunning{Topological descriptors for coherence in CBL}
%
\author{Jos\'e Lic\'on-Sal\'aiz\inst{1}\orcidID{0000-0002-8733-2256} \and
    Cedrick Ansorge\inst{2}\orcidID{0000-0001-9913-3759}}
\authorrunning{J. Lic\'on-Sal\'aiz and C. Ansorge}
%
\institute{Mathematical Institute, University of Cologne, Weyertal 86-90, 50931 Cologne, Germany \and
Institute of Geophysics and Meteorology, University of Cologne, Pohligstr. 3, 50969 Cologne, Germany
\\
\email{licon@math.uni-koeln.de}}
\maketitle              
\begin{abstract}
The interaction between a turbulent convective boundary layer (CBL) and the underlying
land surface is an important research problem in the geosciences. In order to model this
interaction adequately, it is necessary to develop tools which can describe it quantitatively. Commonly
employed methods, such as bulk flow statistics, are known to be insufficient for this task,
especially when land surfaces with equal aggregate statistics but different spatial patterns
are involved.
While geometrical properties of the surface forcing have a strong influence
on flow structure, it is precisely those properties that get neglected when computing bulk statistics.
Here, we present a set of descriptors based on low-level topological information (i.\,e. connectivity),
and show how these can be used both in the structural analysis of the CBL and in modeling its response
to differences in surface forcing. The topological property of connectivity is not only easier to compute
than its higher-dimensional homological counterparts, but also has a natural relation to the physical
concept of a coherent structure.

\end{abstract}
\section{Introduction}\label{sec:introduction}

Thermal convection is a classical pattern-forming physical system, which in idealized settings is known
to produce hexagonal cell structures~\cite{Cerisier1996}. This type of pattern formation exists beyond simple,
idealized conditions and appears also in geophysical settings, where it is a manifestation of
self-organization in non-linear dynamical systems perturbed away from equilibrium~\cite{Goehring2013,Vereecken2016}.

We study here a convective boundary layer (CBL, representing the lowest
portion of the earth atmosphere) which is energetically forced by solar heating of the land surface.
This energy input initiates and sustains convection, and if it is sufficiently strong,
turbulence arises due to natural flow instabilities. This destroys the simple geometrical patterns found
in the ideal Rayleigh-B\'enard case and gives rise to a new type of organized motion where large-scale structures
emerge, such as convective plumes. These are connected regions of the spacetime domain with positive buoyancy
and can thus be seen as a type of coherent structure.
Furthermore, interaction with the land surface modulates these processes, affecting such physical characteristics
as the time scales of land-atmosphere coupling~\cite{Shao2013} and the cloud size
distribution~\cite{Rieck2014}.
The role of geometrical properties of the forcing pattern in determining the structure of the flow has
recently come under study as well~\cite{Kondrashov2016}. 

Descriptions of a CBL usually employ global statistical measurements, such as bulk profiles or spectral
transform of model variables~\cite{Kaimal1994}. The limitations of this approach are encountered, for example,
when using such descriptors to discriminate between the effects of land surface patterns of varying
heterogeneity on the structure of CBL flow~\cite{Liu2017}. Here, we present a methodology that relies on the
construction of a geometrical representation of flow structures, and uses its connectivity information as
descriptor for the state and structure of the CBL. We show how this type of information
corresponds to classical results on scaling properties of turbulent flows, and how it can be effectively used
as a set of features in a statistical learning framework to perform inference on the land-atmosphere
interaction problem. Furthermore, the spatial relationships encoded by connectivity are also found to describe
the existence and structure of the \textit{plume-merging layer} within the CBL~\cite{Mellado2016}.

Previous applications of topology to the study of non-linear physical systems have focused on establishing the
existence of chaotic behavior and characterizing the response of such systems to noise~\cite{Gameiro2004,Gameiro2005};
describing the structural characteristics of transition to fully-developed turbulence in confined plasma
flows~\cite{Garcia2009}; and specifically in the case of convection, the analysis of spiral-defect patterns which arise
after the onset of convection in a low Prandtl number regime~\cite{Krishan2007}. All these studies rely on the
computation of homological invariants in dimensions $0, 1$ and $2$
(the Betti numbers). The present study, in contrast, emphasizes the use of low-level connectivity information
(equivalent to zero-dimensional homology) to furnish physically meaningful descriptors for the state of a CBL in
fully-developed turbulence, and introduces a different connectivity-based topological invariant, the merge tree, to
give a quantitative description of the coherent structures formed therein. Computation of these connectivity-based
descriptors is less computationally expensive than for their higher-order counterparts. Moreover, the connected
components of a domain can be directly related to physical structures in the data. This is not necessarily the case for
homological features of higher dimension, for which the problem of finding a representative cycle that corresponds to a
feature in the data is subject of ongoing research~\cite{Obayashi2018a}.

\section{Methodology and data}

\subsection{Data}
\label{subsec:data}

\begin{figure}[t]
\centering
\includegraphics[width=0.75\linewidth]{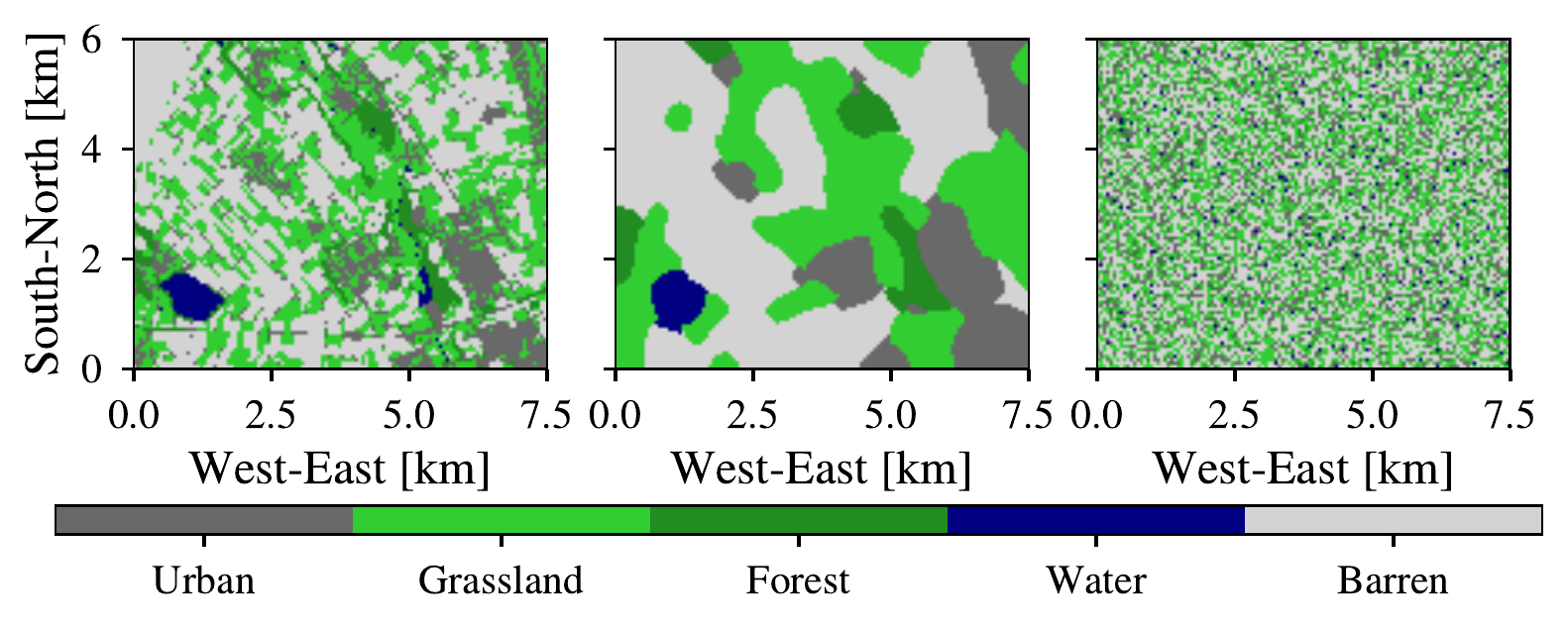}
\caption{Heterogeneous land surface patterns used in the LES-ALM simulations. SP1 (left) shows the real land use;
    SP2 (center) is as SP1 with small-scale features filtered out; SP3 (right) is a randomized pattern. All land
    types have the same relative frequency throughout the three patterns.}
\label{fig:land-surface-types}
\end{figure}

We use data from numerical simulations of a CBL. Specifically, four simulation runs of the Large-Eddy
Simulation Atmosphere--Land-Surface Model (LES-ALM) as described in~\cite{Shao2013}. The four simulations represent a
dry boundary layer on Aug 5, 2009, based on a radio sounding at 08:00h, with the setup as described in~\cite{Liu2017}.
Each of the four simulations, referred to as SP1-4 in the sequel, features different boundary conditions at the
surface. SP1 has the original land use data of the Selhausen-Merken site in Western Germany; SP2 has the same relative
frequency of each land type as SP1, but all small-scale features are filtered out; SP3 is a fully randomized pattern
with the relative frequency for each type; SP4 is a homogeneous grassland surface. See
Fig.~\ref{fig:land-surface-types} for an illustration of the three heterogeneous
land surface patterns. The computational domain covers an area of $7.5 \times 6.0$ \si{\kilo\metre\squared} in the
horizontal, with grid spacing of $\Delta x = \varDelta y = 60$ \si{\metre}, and a height of \SI{2.2}{\kilo\metre}, with
variable spacing of up to $\Delta z = $ \SI{24}{\metre}. Model time step is \SI{0.2}{\second}, with data saved for
analysis at \SI{1}{\minute} intervals. Simulation time is 0800--2000 UTC, 5 Aug 2009. Lateral boundary conditions are
periodic.

\subsection{Methodology}
\label{subsec:methodology}

We will assume a computational domain  $\Omega \subset \mathbb N^4$, with
\[
    \Omega = \{ 1, \ldots, N_t \} \times \{ 1, \ldots, N_z \} \times \{ 1, \ldots, N_y \} \times \{ 1, \ldots, N_x \}.
\]
Each number $N_\bullet$ denotes the number of grid points in each of the four dimensions (time, height, North-South, and
East-West, in that order).
For the present study we will focus on the vertical wind velocity, $w$, as it is a direct representation of the
exchange of energy and momentum at the heart of a convective
system. The method then consists of two steps: the construction of a geometric representation
of the flow, and computation of its connectivity information.
%
%
%
\subsection{Geometric representation}\label{subsec:geometric-representation}
We will use two- and three-dimensional \textit{cubical complexes} as geometrical representations of flow structures
(for background on cubical complexes and their homology, see~\cite{Kaczynski2004}).

\begin{figure}[t]
    \centering
    \includegraphics[width=\linewidth]{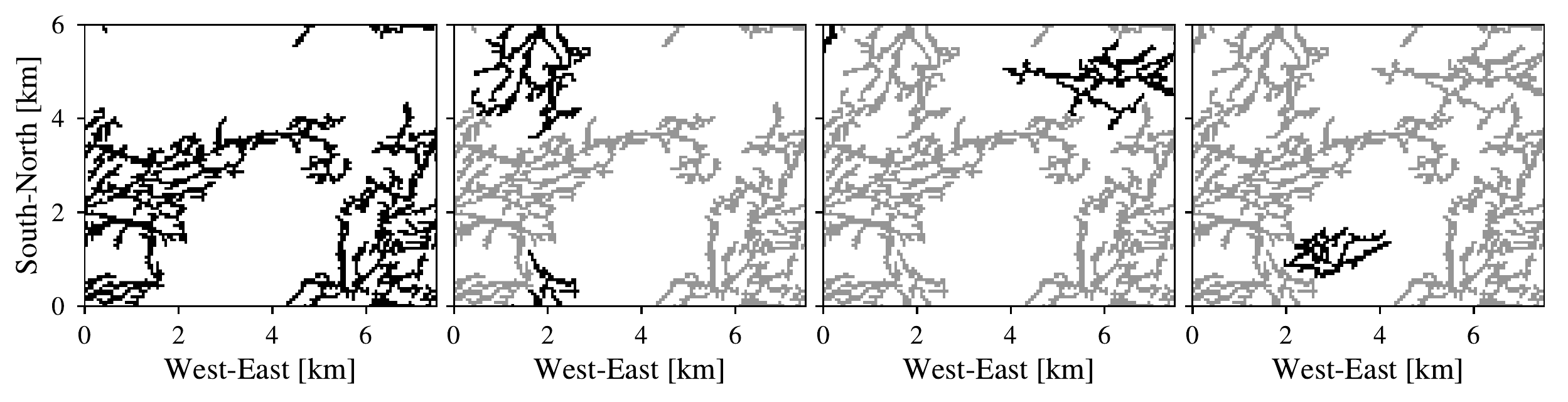}
    \caption{The four largest connected components of the updraft domain $\mathcal C^+$ in a two-dimensional cross section of
        LES-ALM simulation SP1, at \SI{13}{\hour}, \SI{44}{\metre} height.
        The first panel shows the largest connected component with an area of \SI{7.88}{\square\kilo\metre}.
        The next panels show the second, third, and fourth largest components (in black), each having sizes of
        \SI{1.96}{\square\kilo\meter}, \SI{1.5}{\square\kilo\metre}, and \SI{0.82}{\square\kilo\metre} respectively.}
    \label{fig:largest-components}
\end{figure}

Let $t \in \{ 1, \ldots, N_t \}$ and $z \in \{ 1, \ldots, N_z \}$. We denote by $\Omega_t$ the subset of the domain $\Omega$
where the time coordinate is equal to $t$ (this subset is then a three-dimensional array of points). We define $\Omega_{t, z}$ analogously,
as the subset of $\Omega$ with time equal to $t$ and height equal to $z$. Given a threshold value $\varepsilon > 0$, we define
the following sets of points:
\begin{equation*}
    \mathcal P_{t, z}^+ = \{ (k, l) \in \Omega_{t, z} \mid w(t, z, k, l) > \varepsilon \},
\end{equation*}
\begin{equation*}
    \mathcal P_{t}^+ = \{ (j, k, l) \in \Omega_t \mid w(t, j, k, l) > \varepsilon \}.
\end{equation*}
We will use these as \textit{anchor points}
in the construction of the following two- and three-dimensional cubical complexes:
\begin{equation*}
    \mathcal C_{t, z}^+ = \bigcup_{(k, l) \in \mathcal P_{t, z}} [k, k + 1] \times [l, l + 1] \subset \mathbb R^2,
\end{equation*}
\begin{equation*}
    \mathcal C_{t}^+ = \bigcup_{(j, k, l) \in \mathcal P_{t}} [j, j + 1] \times [k, k  + 1] \times [l, l + 1] \subset \mathbb R^3.
\end{equation*}
These geometrical objects will represent the area and volume of the subdomains which correspond to updrafts, as shown
in Fig.~\ref{fig:largest-components}. When it is clear from context we omit the subindices $t,z$ and refer to these
as simply $\mathcal{C}^+$. We will also refer to the first two Betti numbers of the two-dimensional complexes $\mathcal C_{t,z}^+$
as $\beta_0^+$ and $\beta_1^+$, which count the number of connected components and the number of loops, respectively.

\subsection{Two-dimensional analysis}
\label{subsec:2d-analysis}

\begin{figure}[t]
    \centering
    \includegraphics[width=1\linewidth]{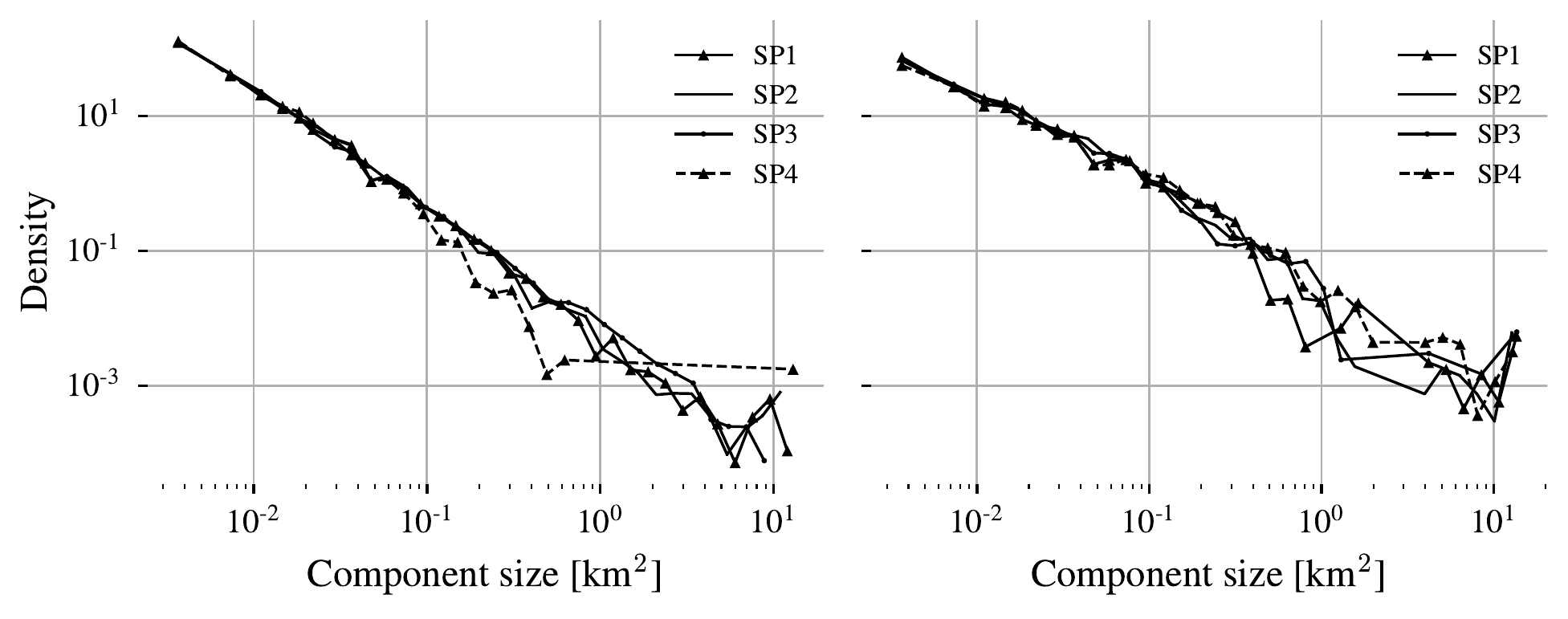}
    \caption{Empirical PDF of the variable $p_j$, the fraction of connected components of size $j$,
        in the positive vertical wind velocity domains for a two-dimensional cross section.
        Both panels show data for the 30 timesteps between 13:00 h and 13:30 h. 
        Left: \SI{44}{\metre}, right: \SI{1100}{\metre}.}
    \label{fig:size-histograms}
\end{figure}

Consider a two-dimensional cubical complex $\mathcal C^+ = \mathcal C_{t,z}^+$, as shown in Fig.\ref{fig:largest-components}.
We note that the sizes of its connected components
are not distributed uniformly at random. Instead, they follow a specific pattern characterized by the existence of
one large component, which accumulates most of the domain area, and a large number of much smaller components.
Denote the connected components of $\mathcal C^+$ by $c_i$, the size of a component by $j$, and the fraction of 
components with size equal to $j$ by $p_j$. Observe that, even if we express the size $j$ in \si{\kilo\metre\squared},
it is still a \textit{discrete} quantity. Finding the connected components of an object such as
$\mathcal C^+$ is a classical problem in computer science,
which has a very efficient algorithmic solution as described by~\cite{Hopcroft73}, making use of the Union-Find (UF)
data structure. For the present study we implement this solution as described in~\cite{Edelsbrunner2010},
which requires storing the size of each component as a criterion to assign precedence when merging two components
together. Thus, we get the size information essentially ``for free'' when splitting $\mathcal C^+$ into its components.
The empirical distribution of $p_j$ is shown in Fig.~\ref{fig:size-histograms},
for the cubical complexes obtained from two-dimensional slices with $z = 44$ \si{\metre} (left) an
$z = 1100$ \si{\metre} (right). In both cases the data for 30 timesteps is shown, that is, the sizes
of components from 30 different cubical complexes are aggregated.

The general pattern becomes clear in these figures: the smallest sizes accumulate most of the density for $p_j$.
This density then decreases regularly as size increases,
until it is several orders of magnitude smaller for the very largest components. This regular behavior is clearer in the 
data from the surface layer (left panel), and especially in the sizes for simulation SP3 (randomized land surface pattern).
The distribution appears to decay linearly in a log-log scale. There is a marked difference between the density curve for 
SP4 (homogeneous land surface pattern) and SP1-3, but the three heterogeneous surface patterns exhibit a mostly similar curve.
This serves to illustrate an important difference between the homogeneous and heterogeneous cases, namely a change in the dominant
physical mechanism. For SP1-3, heterogeneity in the CBL is maintained as a consequence of forcing over a range of spatial scales,
whereas in SP4 structures are produced only as a consequence of turbulence internal to the boundary layer. In other words, the 
absence of forcing at the largest spatial scales in the case of SP4 becomes manifest here as a scale break at 
$\sim$\SI{0.7}{\square\kilo\metre}.
For the data from the mixing layer (right panel) the first part of the distribution still appears to decay linearly on the log-log scale,
but there are now much larger fluctuations at the tail.

We will use the scaling patterns just described to analyze the effects of the different land surface patterns on
the structure of CBL flow. Specifically, we will fit a parametric heavy-tailed distribution to the data,
and perform the analysis on its scaling parameter. A reasonable choice of distribution is the \textit{power law},
defined in its discrete form by 
\begin{equation}\label{eq:powerlaw}
p(x) = C \, \chi_{[x_{\min}, \infty)} (x) \, x^{- \alpha},
\end{equation}
where $\alpha$ is the scaling parameter and $C = 1 / \zeta(\alpha, x_{\min})$ a normalization constant%
\footnote[1]{$\zeta$ is here the generalized zeta function.}. 
A second parameter 
is the support of $p(x)$, represented by a value $x_{\min}$ which gives the lower bound for the power-law scaling.
Distributions of this form are important in different areas of science~\cite{Mitzenmacher2004}.
A statistical framework based on maximum-likelihood parameter estimation to analyze empirical data and
ascertain whether it conforms to a power-law distribution has been proposed in~\cite{Clauset2009}.
For the results presented in this paper, we have used the Python implementation of these methods contained 
in the package \texttt{powerlaw}~\cite{Alstott2014}.

An important consideration to keep in mind before doing this is the amount of data points available.
Following~\cite{Clauset2009} we take $n = 50$ as the minimum size to obtain reasonable fit results.
However, we find that $\beta_0^+ > 50$ only for a small proportion of all the complexes $\mathcal C_{t,z}^+$.
Denote the range of simulation timesteps by $t = 0, 1, \ldots, N_t$. We begin by defining a time window $t_w$,
and splitting the time range into the time intervals $T_i$ defined as
\begin{align*}
T_1 &= \{ 0, 1, \ldots, t_w - 1 \} \\
T_2 &= \{t_w, t_w + 1, \ldots, 2\,t_w -1 \} \\
&\vdots    \\
T_M &= \{(M-1)\, t_w, \ldots, N_t \}.
\end{align*}
Given a height value $z$ and a time interval $T_i$, we will have $t_w$ two-dimensional slices, except perhaps for the
last time interval, $T_M$. For each of these slices, $\ell \in T_i$, we construct the cubical complex
$\mathcal C_\ell$, This complex is conformed by $n_\ell$ connected components which we denote by
$c_{k, \ell}$, such that
\[
\mathcal C_\ell = \bigcup_{k=1}^{n_\ell} c_{k, \ell}.
\]
Let
$s(c_{k, \ell})$ denote the size of component $c_{k, \ell}$. We then define
$\mathcal S_\ell = \{ s(c_{k, \ell}) \mid c_{k, \ell} \text{ conn. component of } \mathcal C_\ell \}$
as the set of all sizes of connected components that make up $\mathcal C_\ell$. Define furthermore the
aggregate set 
\begin{equation}
\mathcal S_{T_i} = \bigcup_{\ell \in T_i} \mathcal S_\ell,
\end{equation}
which accumulates the sizes of all connected components observed during the time interval $T_i$, at height level $z$.
We then fit a power law distribution to the values $p_j$ for this set. We thus increase the sample size
used to fit the power-law distribution, while allowing the time dependence of the power-law scaling behavior
to be observed. In what follows, we use a time window of $t_w =$ \SI{10}{min},
which results in \num{73} time periods (\(N_t = 721\)). This means we will fit \num{7300} power law distributions
per simulation dataset, obtaining from each distribution its scaling parameter $\alpha$.

\subsection{Three-dimensional analysis}
\label{subsec:3d-analysis}

\begin{figure}[t]
    \centering
    \includegraphics[width=1\linewidth]{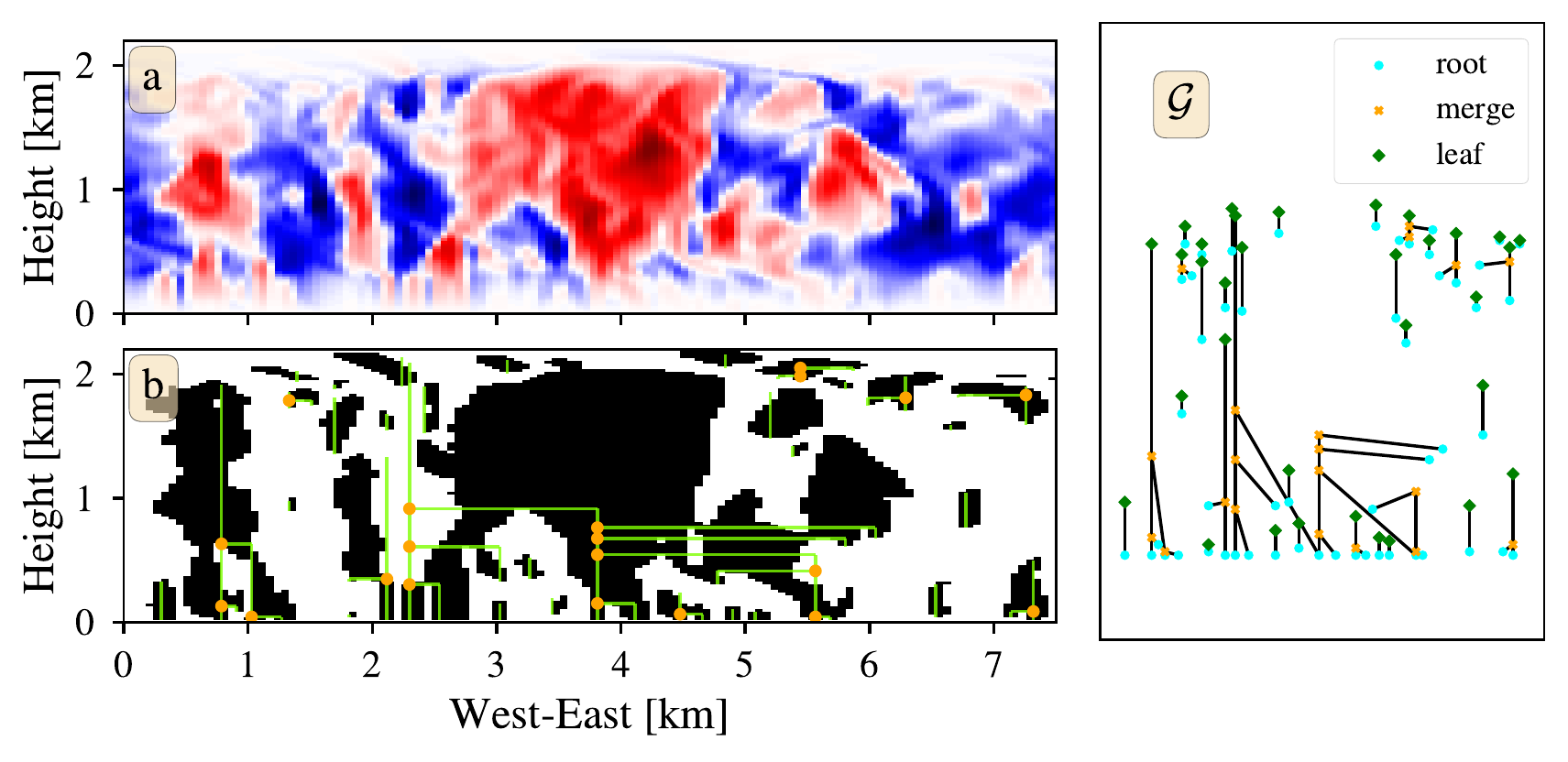}
    \caption{Merge tree obtained for a vertical two-dimensional slice. a) shows the values of vertical wind velocity $w$
        (red: $w>0$, blue: $w<0$). b) the black region is the cubical complex $\mathcal C^+$ obtained by thresholding the
        data, the orange dots indicate merge events. $\mathcal G$ shows the graph representation of the merge tree.}
    \label{fig:merge-tree-example}
\end{figure}

Let $\mathcal C^+ = \mathcal C_{t}^+$ be a three-dimensional cubical complex. Define its \textit{height function}, 
$f : \mathcal C^+ \rightarrow \mathbb N$ by
\begin{equation*}
    f(z, y, x) = j,
\end{equation*}
where $j$ is such that $z \in [j, j+1]$. That is, for every point in the complex, this function returns the height of the three-dimensional
elementary cube that contains it. Define, further, the following equivalence relation on points of $\mathcal C^+$:
$u \sim v$ iff $f(u) = f(v)$ and both $u$ and $v$ belong to the same connected component of the sublevel set
$f^{-1} (-\infty, f(u)]$.
The \textit{merge tree} of $f$ is the quotient space $\mathcal C^+ / \sim$. This topological invariant tracks the evolution
of sublevel sets of $f$, which can be born at level $z$ if a new independent component appears at that level, or die if they
are merged into an older component when entering $z$, or stop growing altogether. An example of this is shown
in Fig.~\ref{fig:merge-tree-example}, where the graph $\mathcal G = \mathcal G_t$ represents the merge tree obtained
from the scalar field $w$. We distinguish three types of nodes:
\textit{root} nodes represent the birth of an independent component, \textit{merge} nodes the death of a component
by merging, and \textit{terminal} nodes represent the level at which a component stops growing in the vertical direction.
It is possible to perform statistical analysis on the set of graphs obtained by this method, but for the results
discussed here we will focus on the spatiotemporal distribution of merge events.

\section{Results and discussion}
\label{sec:results}

\subsection{Updraft size scaling}

\subsubsection{Scaling parameter}
\label{subsec:pl-scaling-parameter}

\begin{figure}[t]
    \centering
    \includegraphics[width=0.45\textwidth]{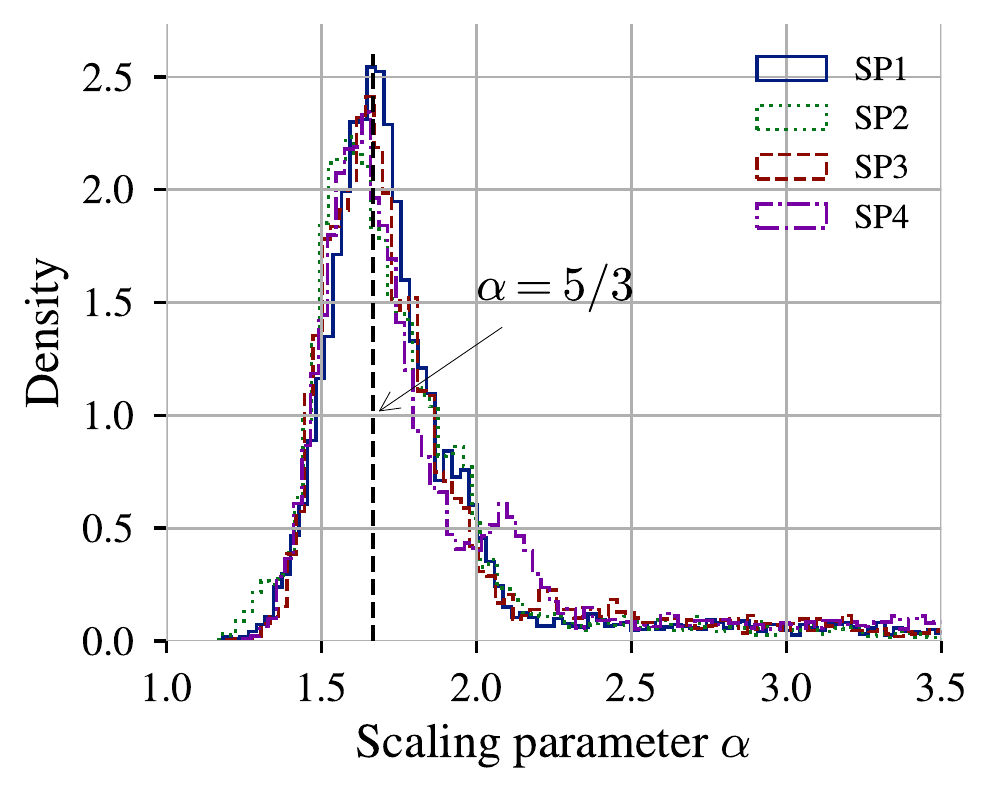} \includegraphics[width=0.5\linewidth]{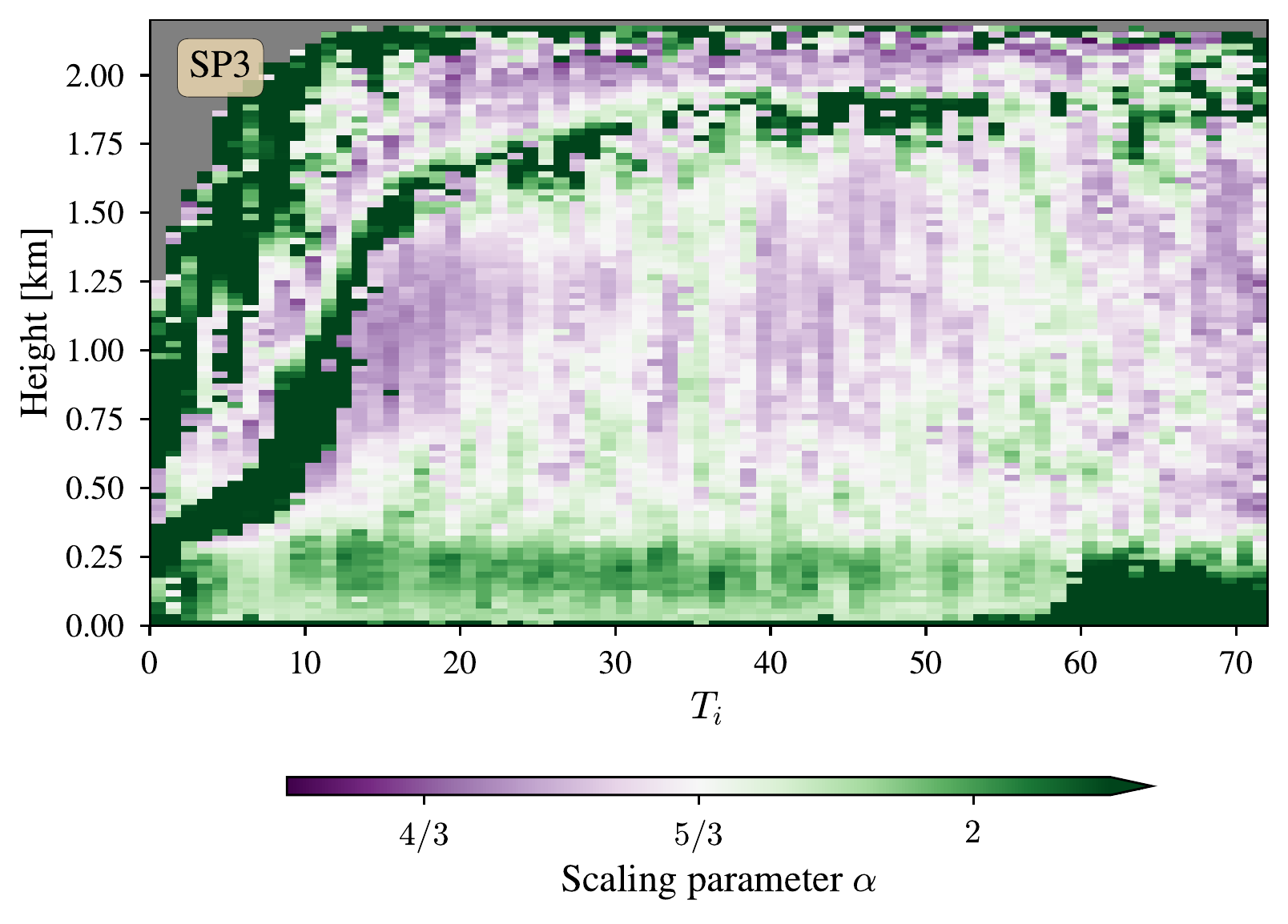}
    \caption{Left: normalized histograms of all power law exponent $\alpha$ values, with a vertical line indicating the value $\alpha=5/3$,
        which corresponds to the Kolmogorov scaling law. Right: distribution of $\alpha$ in the $(t,z)$-plane for simulation SP3,
        showing that the value $5/3$ occurs predominantly in the turbulent region. The distribution for SP1, SP2 and SP4 is very similar.}
    \label{fig:power_law_exponents}
\end{figure}

The empirical PDF of the scaling parameter $\alpha$ for all power law densities fitted for the four LES-ALM
datasets are shown in Fig.~\ref{fig:power_law_exponents} (left). These values are very similar across the
four simulations, with the only significant difference being that the distribution for SP4 has a more
clearly defined bimodal structure. The value distribution appears reasonable, insofar as power law
densities encountered in practice tend to have a scaling parameter within the range
$1 \leq \alpha \leq 2$~\cite{Mitzenmacher2004}. Further, the maximum of the PDF in all four cases occurs
in a narrow band around $\alpha \approx 5/3$, which is the value expected from the self-similarity of
eddies cascading the spectrum of motion~\cite{Kolmogorov1941a,Kolmogorov1941b,Obukhov1941,Kaimal1994}.

The time-height sections of the scaling parameter $\alpha$ are illustrated in Fig.~\ref{fig:power_law_exponents} (right), as a function 
of height and time. The distribution of $\alpha$ in the $(t,z)$-plane is clearly non-uniform, and is seen to reflect the division of the
CBL into its component regions. The qualitative features seen here are:
\begin{itemize}
    \item The existence of a well-defined surface layer, encompassing the first
    \num{15} height levels closest to the land surface.
    \item The evening transition is apparent as a sharp increase in the mean
    value of $\alpha$ close to the surface, starting around $T_i = 57$. This
    effect is most pronounced for SP3 and SP4.
    \item The inversion layer is characterized by larger $\alpha$ values than 
    those found either in the mixing layer below it, or in the free atmosphere above.
    \item Values of $\alpha \approx 5/3$ occur predominantly within the mixing layer, which agrees
    with the fact that this is the region where strong turbulence develops.
\end{itemize}

\subsubsection{Goodness-of-fit}
\label{subsec:goodness-of-fit}

It is difficult to accurately discriminate a power-law distribution from empirical data, due to the
possibility of error introduced by the appearance of very large values at the tail of the distribution, which is one of the defining
characteristics of a power law. We perform the following goodness-of-fit testing on the available data, based on the 
method described in~\cite{Clauset2009}: let $X = \{x_1 , \ldots, x_n\}$
be the data set, and $f(x) = f(x;\, \hat \alpha, \hat x_{\min})$ the power law density fitted to it, with corresponding CDF denoted by
$F(x)$. The empirical CDF for the data is denoted by \(G(x)\). The method then computes the Kolmogorov-Smirnov (KS) statistic
for \(F\) and \(G\), and compares it to that obtained for \(n\) synthetic datasets. The fraction of these for which their
KS statistic is larger than that obtained for the data is the p-value for this test.
Within this framework, it is small p-values that signal insufficient empirical evidence in favor of the power-law hypothesis.
Following~\cite{Clauset2009} we take ``small enough'' to be $p < 0.1$. We use a value of $n = 2500$, which would give a p-value accurate
up to 2 decimal places~\cite{Clauset2007}. Using these parameters we find that the hypothesis of power-law scaling is strongest
within the surface layer, and in isolated mixing layer points. In terms of the four land-surface patterns, the hypothesis is strongest
for SP3, the randomized surface pattern, and weakest for the surface layer in SP4. Overall this is in agreement with the data shown
in Fig.~\ref{fig:size-histograms}, where the scaling for the data from surface layer in case SP3 is the most regular of all four simulations.
In all four cases, this scaling behavior becomes less frequent after the evening transition takes place ($T_i = 57$).
\subsubsection{Likelihood ratio test}
\label{subsec:likelihood-ratio}
We compare the likelihood of the data under the power-law hypothesis, $\mathcal L_0 = \mathcal L_0 (X)$, with the
likelihood under an alternative heavy-tailed distribution, $\mathcal L_1 = \mathcal L_1(X)$.
Common alternatives are the exponential and the log-normal. We then compute the log-ratio
\begin{equation}\label{eq:loglikelihood-ratio}
R = \log \left( \frac{\mathcal L_0}{\mathcal L_1} \right),
\end{equation}
and an associated p-value for the likelihood ratio test~\cite{Vuong1989}. In this case, ``small'' p-values would indicate that
the observed value is unlikely to be the result of random fluctuations alone. If the p-value is larger than a certain threshold,
we consider the data to be insufficient for this test. Otherwise, the sign of $R$ indicates the hypothesis favored by the data.

Comparing the power-law hypothesis to an exponential alternative shows that the latter is not favored by the data.
Indeed, the p-values for the corresponding log-likelihood test (not shown here) are smaller than \num{0.01} for \num{79}\% of the $(z, T_i)$ 
pairs from SP1, \num{82}\% for SP2, \num{78}\% for SP3, and \num{76}\% for SP4. Moreover, the points where this happens agree with
those where the log-likelihood ratio is positive and greatest in magnitude.

The log-likelihood test for a power law against a log-normal distribution paints a more nuanced picture.
At a significance level of \num{0.05}, the test is actually not able to distinguish between either of the two hypothesis 
for the vast majority of the domain. More important for our current analysis is that most of the surface layer points, for which
the power law hypothesis has the greater likelihood, fall within the statistically significant tests (at the \num{0.05} level).
We summarize the results of this subsection so far:
\begin{enumerate}
    \item The original observation made at the beginning, that the logarithm of frequency at which connected updraft regions
    appear in two-dimensional slices of the LES-ALM simulations decreases linearly with the logarithm of their size, finds limited
    evidential support across the totality of simulation data.
    \item There is a strong indication of power-law scaling for the size of connected updraft components within the surface layer,
    and the degree to which a power-law distribution is a good fit for the data appears to be in direct relation to the heterogeneity
    of the underlying land surface. In all four simulations, the scaling parameter $\alpha$ for the surface layer data
    is, on average, in the range $1.9 \leq \alpha \leq 2.0$.
    \item Comparison of the power-law distribution with the alternative of an exponential or a log-normal distribution via a log-likelihood
    ratio test shows that, for the surface layer, the power law is in general the most adequate to describe the distribution of connected
    component size.
    \item The evening transition brings an end to the power-law scaling behavior in the surface layer. After this point in time, the power-law
    distribution appears only in one vertical level adjacent to the land surface for SP3.
\end{enumerate}

\subsubsection{Comparison with the updraft Betti number, $\beta_0^+$}
\label{subsec:power-law-betti-comparison}

\begin{figure}[t]
    \centering
    \includegraphics[width=\linewidth]{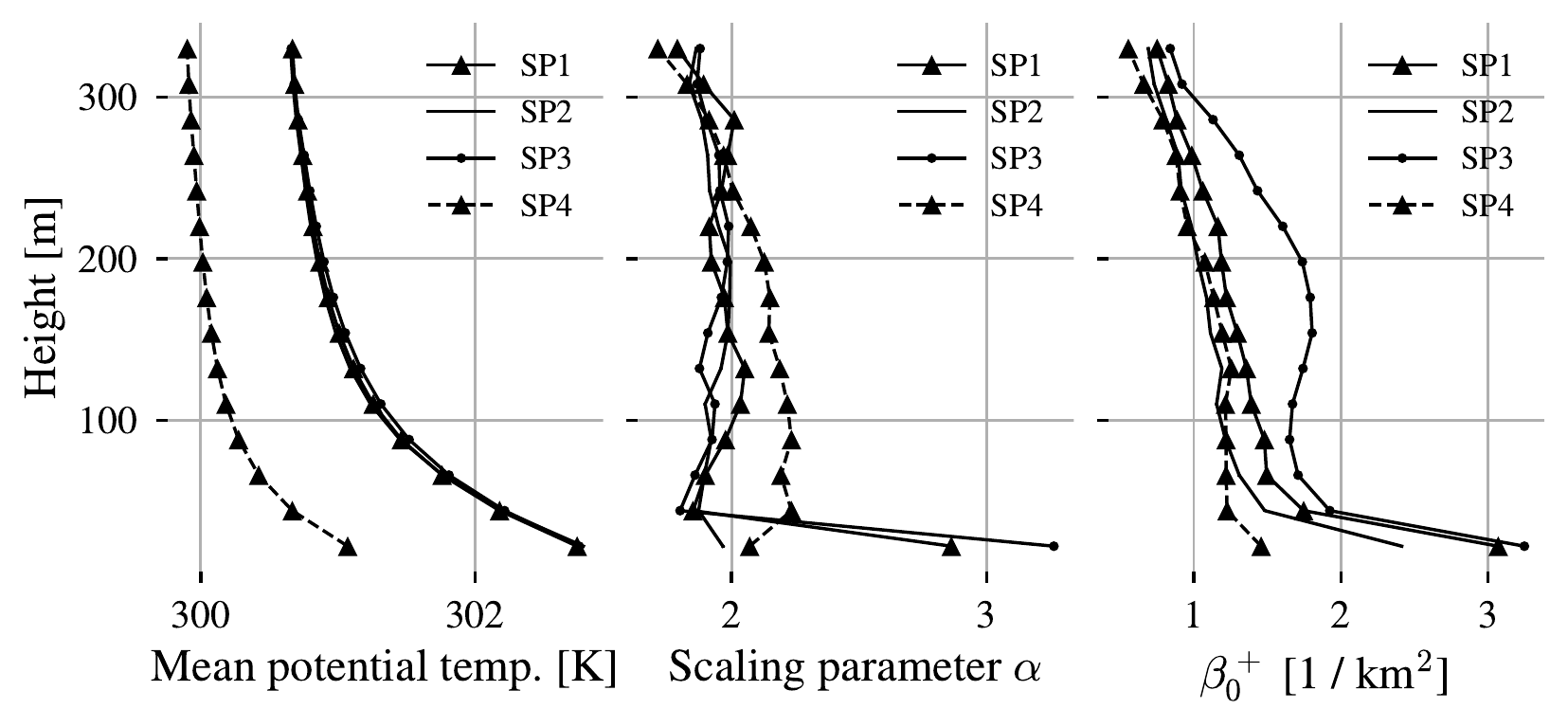}
    \caption{Vertical profiles of mean potential temperature T (left), power-law scaling parameter
        $\alpha$ (middle) and zeroth Betti number $\beta_0^+$ (right) for the first 15 vertical levels,
        corresponding to the surface layer. The values of $\alpha$ correspond to time interval $T_i = 40$.
        For T and $\beta_0^+$, the \num{10} \SI{1}{\minute} intervals are averaged.}
    \label{fig:profile_comparison}
\end{figure}

We now turn to the question of recovering the land surface pattern using only the information provided by both the scaling parameter
$\alpha$ and the zeroth Betti number $\beta_0^+$. In light of the discussion above, we will focus on the data emanating from the surface
layer. A comparison of these values for the 4 LES-ALM simulation runs is shown in
Fig.~\ref{fig:profile_comparison}, together with the mean temperature profile. As was discussed before, it is known that the
temperature profile can in general discriminate between homogeneous and heterogeneous land surface patterns, but fails at reflecting
the effect of different heterogeneous patterns. Both $\alpha$ and $\beta_0^+$ seem better
at expressing this difference.

To make this more precise, we translate the aforementioned question to a classification problem. For each simulation SP$k$
($k = 1,\ldots,4$) consider the matrix $X_{\alpha, k}$, formed by \num{60} observations (rows) and \num{15} features (columns).
The entry $X_{\alpha,k} [i, j]$ will be equal to the value of $\alpha$ obtained from the data at time window $T_i$, height level $j$,
for simulation SP$k$. Since we restrict the data to the surface layer, $j = 1, \ldots, 15$.
We then construct the feature matrix $X_\alpha$ by concatenation:
\[
\renewcommand\arraystretch{1.3}
X_\alpha = 
\mleft[
\begin{array}{c}
X_{\alpha, 1}  \\
\hline
X_{\alpha, 2}  \\
\hline
X_{\alpha, 3}  \\
\hline
X_{\alpha, 4}  \\
\end{array}
\mright],
\quad
y =
\mleft[
\begin{array}{c}
\text{SP}1 \\
\hline
\text{SP}2 \\
\hline
\text{SP}3 \\
\hline
\text{SP}4 \\
\end{array}
\mright].
\]
The response variable $y$ is simply the label SP$k$ for each of the \num{240} observations. We construct the feature matrix
$X_\beta$ for the zeroth Betti number in analogous fashion, where the entry $X_{\beta,k} [i, j]$ now contains the mean value
of $\beta_0^+$, averaged over the individual timesteps within $T_i$, at height level $j$, for simulation SP$k$. For comparison
we use the temperature data as $X_{\theta,k}$.
A k-nearest-neighbors (kNN) classifier is trained on each (mean-centered and rescaled to unit variance) feature matrix
separately, and then evaluated by computing the
bootstrap estimate of the weighted average $F_1$ score ($n = 1000$ samples). This estimate was computed for values of
$k = 1, \ldots, 15$. For $X_{\alpha}$ and $X_{\beta}$ best performance was achieved with $k = 3$, 
whereas for $X_\theta$ it was for $k=7$. The results for this are shown in Table~\ref{tab:compare-classifiers}.
The difference between the classifiers for $X_\alpha$ and $X_\beta$ is significant, with an average $F_1$ score for $X_\alpha$ of \num{0.59}, compared to \num{0.80} for $X_\beta$. This
shows that the scaling law found to describe the size of connected updraft components is sensitive to the differences in land surface patterns.
It is not, however, more sensitive to these differences than the \textit{number} of connected components, $\beta_0^+$. In other words, when
trying to determine which surface pattern produced a given set of surface-layer values, knowing how
many connected components are present in the two dimensional slices is more informative than knowing how their sizes
scale.
\begin{figure}[t]
    \centering
    \includegraphics[width=\linewidth]{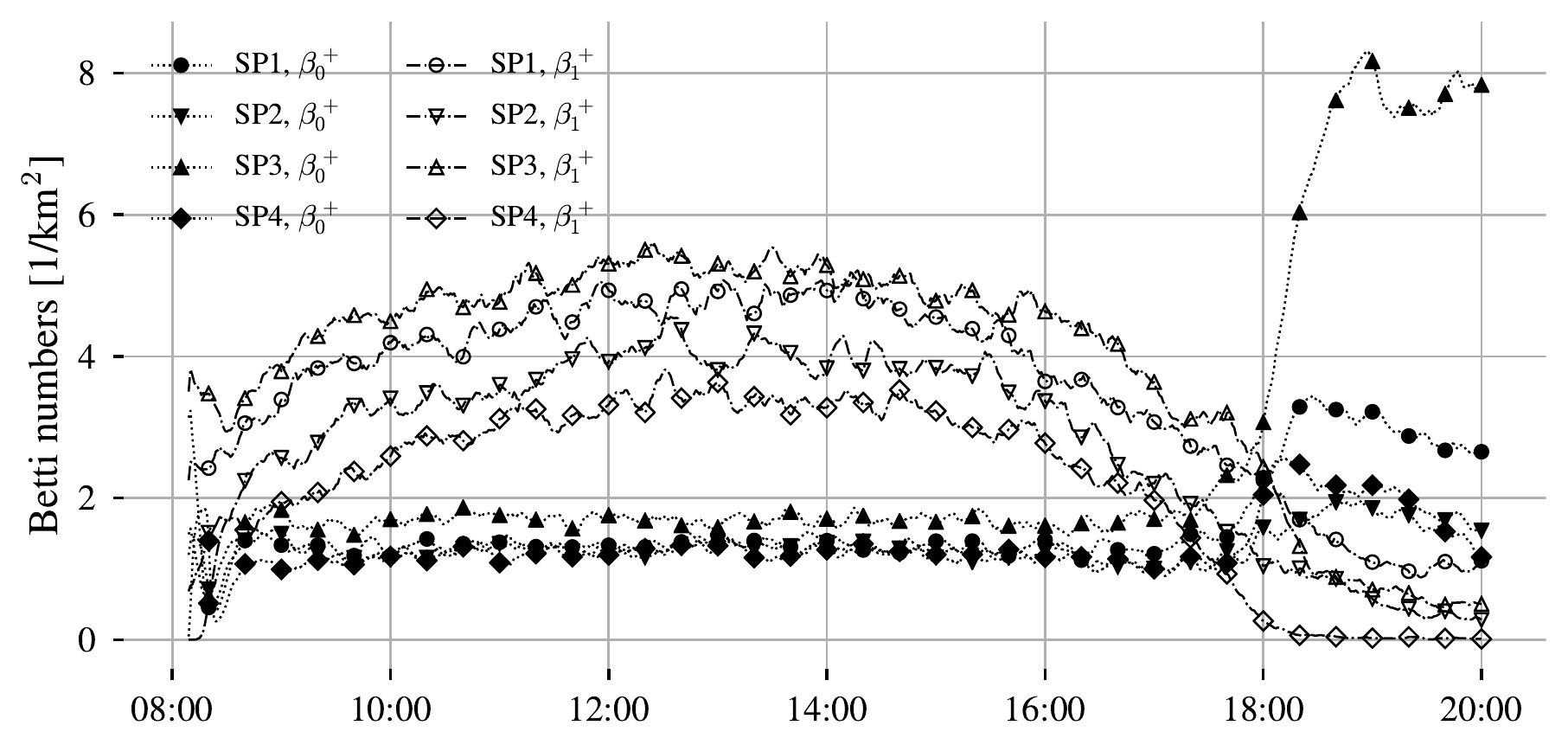}
    \caption{Betti number time series.}
    \label{fig:surface-layer-betti}
\end{figure}
\begin{table}[t]
    \normalsize
    \caption{Performance of k-NN classifiers, $k = 3$, trained with feature matrices using the
        power law exponent $\alpha$ (left) and the zeroth Betti number $\beta_0^+$ (center). The temperature data
        (right) was used to train a k-NN classifier with $k=7$.}
    \label{tab:compare-classifiers}
    \centering
    \begin{tabular}{c|ccc|ccc|ccc}
        \hline
                    &  \multicolumn{3}{c|}{$\alpha$} & \multicolumn{3}{c|}{$\beta_0^+$}  &
                        \multicolumn{3}{c}{$\theta$}\\
        Simulation &  Precision & Recall & $F_1$  & Precision & Recall & $F_1$ & Precision & Recall & $F_1$ \\
        \hline
        SP1 & 0.42 &  0.45 &  0.43 & 0.72 & 0.74 & 0.72 & 0.18 & 0.22 & 0.19 \\
        SP2 & 0.53 &  0.52 &  0.52 & 0.63 & 0.68 & 0.65 & 0.18 & 0.19 & 0.18 \\
        SP3 & 0.60 &  0.61 &  0.60 & 0.97 & 0.96 & 0.96 & 0.16 & 0.14 & 0.14 \\
        SP4 & 0.83 &  0.77 &  0.80 & 0.89 & 0.81 & 0.85 & 0.88 & 0.80 & 0.83 \\
        \hline
        avg. & 0.60 & 0.58 & 0.59 & 0.81 & 0.79 & 0.80 & 0.35 & 0.33 & 0.34 \\
        \hline
    \end{tabular}
\end{table}

The classification metrics shown in the table also show that both sets of features have greatest discriminatory power
when it comes to classifying the SP3 and SP4 data, with the power law model having greatest $F_1$ score for SP4, and the
Betti number model for SP3. Both facts are consistent with the differences in power-law scaling across
the four simulations, to wit: the power law densities fit to size
data can distinguish between the homogeneous land surface pattern SP4 and the three heterogeneous patterns, but 
do not distinguish between the latter three. They are also consistent with the fact
that, throughout most of the day, the value of $\beta_0^+$ for the surface layer of simulation SP3 is on average higher
than it is for any of the other three simulations (see Fig.~\ref{fig:surface-layer-betti}). An interpretation of
these facts would be that land surface heterogeneity introduces an element of scale invariance to the size distribution
of connected updraft regions, with this being clearest for the maximally heterogeneous SP3. Conversely, a purely
uniform land surface, SP4, breaks away from this behavior, thus making its classification easier when only the
scaling parameter $\alpha$ is known.

\subsection{Merge tree representation}
\label{subsec:merge-tree}

\begin{figure}[t]
    \centering
    \includegraphics[width=0.75\linewidth]{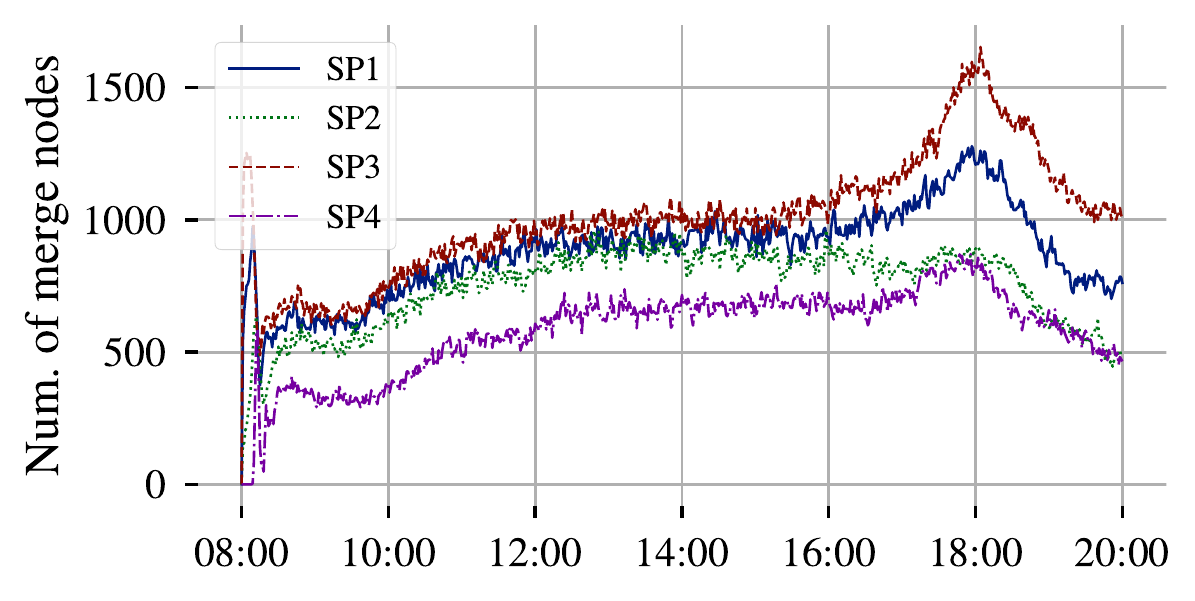}
    \caption{Number of merge nodes at each simulation timestep. The four LES-ALM simulations are shown.}
    \label{fig:number-of-merges}
\end{figure}

We compute the merge tree $\mathcal G_t$ at each timestep $t$, for the four simulations, storing the three
types of nodes that occur (see Fig.~\ref{fig:merge-tree-example}), together with their time and height
indices $(t,z)$. This allows us to visualize the number of merge events in each tree as a time series in
Fig.~\ref{fig:number-of-merges}. It can be seen that the four time series exhibit a similar qualitative
pattern: a brief initial transient, followed by a period of growth until noon, after which the value
stabilizes. There is a second growth phase in the late afternoon, which intensifies at the evening
transition, with a marked decline afterwards. Also noteworthy is the fact that the average values of the four
series are different, and stand in an inverse relationship with land surface heterogeneity length scale. That
is, SP3 has on average the largest number of merge nodes, with SP4 having the smallest. Comparing these time
series with those for $\beta_0^+$ (cf. Fig~\ref{fig:surface-layer-betti}) shows that the number of merge
nodes becomes stationary at a later time than does $\beta_0^+$, and the former reaches its peak after
the evening transition at an earlier time than the latter. Thus, despite the close relationship between both
quantities (in general we can expect more merge events to occur if there are more components to be merged),
they appear to respond differently to land-surface heterogeneity.

\begin{figure}[t]
    \centering
    \includegraphics[width=1\linewidth]{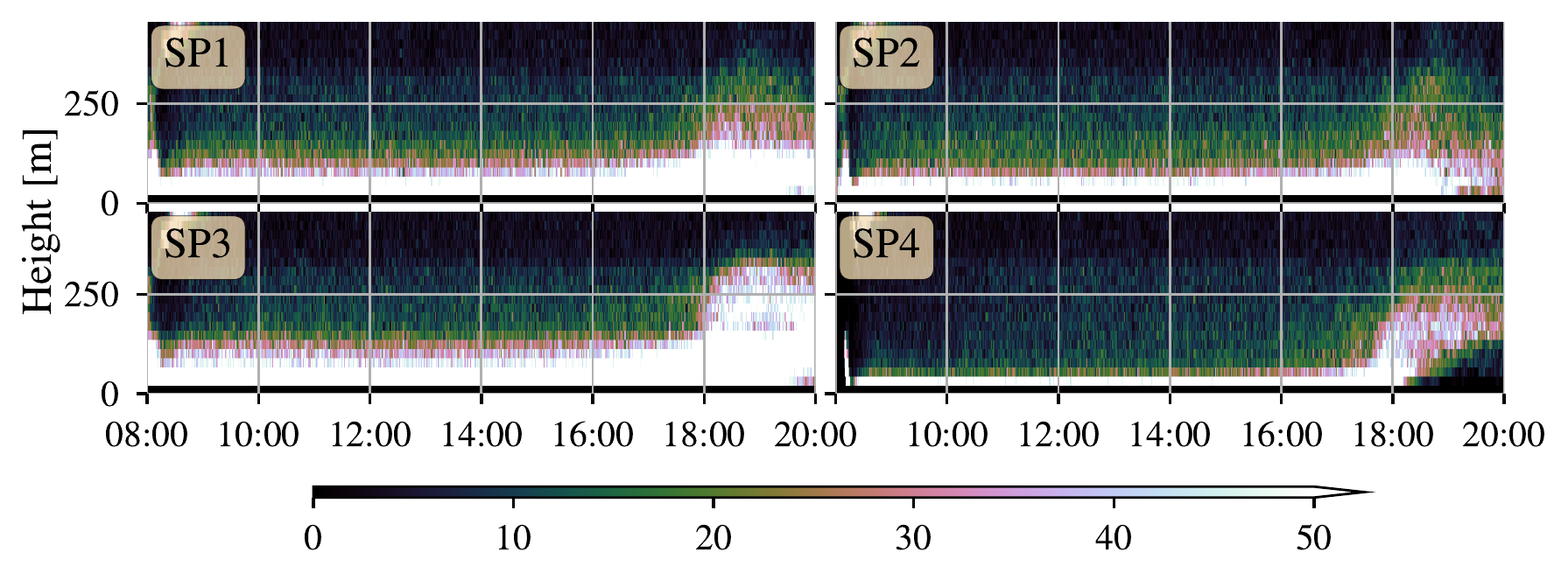}
    \caption{Number of merge nodes for each vertical level within the surface layer, for the four LES-ALM
        simulations.}
    \label{fig:merges-contour-surface}
\end{figure}

A closer view of the surface layer, where most of the merge nodes are concentrated, is given in
Fig.~\ref{fig:merges-contour-surface}. The clearest difference is again that between SP3 and SP4,
with the former having most of its surface layer merge nodes distributed throughout the first 5
vertical levels. In the case of SP4, the merge nodes are concentrated in the first level adjacent
to the land surface. Recall the size distribution shown in \ref{fig:size-histograms}, and the discussion
surrounding the KS goodness-of-fit test for these size distributions. It was shown that the surface layer 
size distributions in SP4
have the worst power law fit out of all four simulations. This fact is expressed in the empirical PDF 
(Fig.~\ref{fig:size-histograms}, left) for SP4 having a scale break followed by a gap of one order
of magnitude, reflecting the presence of the dominant component. What the distribution of merge nodes
suggests is that, in SP4, this dominant component appears already at the first level, whereas for
simulations with increasing levels of land surface heterogeneity the process of component merging
occupies a larger portion of the vertical direction. This shows how the vertical depth of the
plume-merging layer can be modulated by land-surface heterogeneity.

This merging process results in the formation of a large connected structure which accumulates most of
updraft volume in $\Omega_t$, close to 99\%. This also happens very rapidly, within the first \num{30}
simulation minutes in all four cases, and once the structure forms it persists for the remainder of
the simulation.

\section{Conclusions}\label{sec:conclusions}
We have presented an application of topology to the analysis of numerical simulation data from atmospheric
science. Our results specifically highlight the importance of connectivity and its associated topological invariants
(the zeroth Betti number, $\beta_0^+$, and the merge tree) in producing a quantitative description of the
structural properties of a CBL, as well as the response of the CBL system to differences in the underlying
land surface pattern.

Determining the size of connected components in subsets of the computational domain allows us to establish
parametric densities which describe their scaling. The scaling laws thus established are found to correspond
to the expected self-similar scaling in the energy cascade across the spectrum of turbulent motion. Further,
the existence of different subregions that make up the CBL can be inferred from variability in scaling 
across the time and height dimensions.

The number of connected components, $\beta_0^+$, is shown to better discriminate between different
heterogeneous land surface patterns when considering the data close to the land surface. Indeed, we
demonstrate that both $\beta_0^+$ and the scaling parameter $\alpha$ are shown to be more informative in
this problem than the bulk temperature profile.

Computing a merge tree representation for the flow allows us to quantify the rate of plume coalescence, as
well as the spatiotemporal evolution of this process across the CBL diurnal cycle. Using this we can quantify
the effect of land surface heterogeneity on the depth of the plume-merging layer. Component merging is also 
shown to respond differently to land surface heterogeneity than either the scaling parameter $\alpha$ or the 
number of connected components.

This study thus shows these topologically-motivated descriptors as a viable alternative to bulk
profiles and spectral transformations for analyzing data from numerical simulations of a CBL.

\section*{Acknowledgements}
We gratefully acknowledge support by the DFG transregional research collaborative TR32 on
Patterns in Soil--Vegetation--Atmosphere Systems.

%
%
%
\bibliographystyle{splncs04}
\bibliography{cbl_organization}

\begin{thebibliography}{10}
\providecommand{\url}[1]{\texttt{#1}}
\providecommand{\urlprefix}{URL }
\providecommand{\doi}[1]{https://doi.org/#1}

\bibitem{Alstott2014}
Alstott, J., Bullmore, E., Plenz, D.: {Powerlaw: A python package for analysis
  of heavy-tailed distributions}. PLoS ONE  \textbf{9}(1) (2014).
  \doi{10.1371/journal.pone.0085777}

\bibitem{Cerisier1996}
Cerisier, P., Rahal, S., Rivier, N.: {Topological correlations in
  B{\'{e}}nard-Marangoni convective structures}. Physical Review E -
  Statistical, Nonlinear, and Soft Matter Physics  \textbf{54}(5),  5086--5094
  (1996)

\bibitem{Clauset2009}
Clauset, A., Shalizi, C.R., Newman, M.E.J.: {Power-law distributions in
  empirical data}. SIAM Review  \textbf{51}(4),  661--703 (2009).
  \doi{10.1137/070710111}

\bibitem{Clauset2007}
Clauset, A., Young, M., Gleditsch, K.S.: {On the Frequency of Severe Terrorist
  Events}. Journal of Conflict Resolution  \textbf{51}(1),  58--87 (2007)

\bibitem{Edelsbrunner2010}
Edelsbrunner, H., Harer, J.: {Computational Topology: An Introduction}.
  American Mathematical Society (2010)

\bibitem{Gameiro2004}
Gameiro, M., Mischaikow, K., Kalies, W.: {Topological characterization of
  spatial-temporal chaos}. Physical Review E - Statistical, Nonlinear, and Soft
  Matter Physics  \textbf{70}(3 2),  9--12 (2004).
  \doi{10.1103/PhysRevE.70.035203}

\bibitem{Gameiro2005}
Gameiro, M., Mischaikow, K., Wanner, T.: {Evolution of pattern complexity in
  the Cahn-Hilliard theory of phase separation}. Acta Materialia
  \textbf{53}(3),  693--704 (2005). \doi{10.1016/j.actamat.2004.10.022}

\bibitem{Garcia2009}
Garcia, L., Carreras, B.A., Llerena, I., Calvo, I.: {Topological
  characterization of the transition from laminar regime to fully developed
  turbulence in the resistive pressure-gradient-driven turbulence model}.
  Physical Review E - Statistical, Nonlinear, and Soft Matter Physics
  \textbf{80}(4),  1--11 (2009). \doi{10.1103/PhysRevE.80.046410}

\bibitem{Goehring2013}
Goehring, L.: {Pattern formation in the geosciences}. Philosophical
  Transactions of the Royal Society of London A: Mathematical, Physical and
  Engineering Sciences  \textbf{371}(2004) (2013). \doi{10.1098/rsta.2012.0352}

\bibitem{Hopcroft73}
Hopcroft, J.E., Ullman, J.D.: {Set merging algorithms}. SIAM Journal on
  Computing  \textbf{2}(4),  294--303 (1973)

\bibitem{Kaczynski2004}
Kaczynski, T., Mischaikow, K., Mrozek, M.: {Computational Homology}.
  Springer-Verlag New York, 1 edn. (2004)

\bibitem{Kaimal1994}
Kaimal, J.C., Finnigan, J.J.: {Atmospheric Boundary Layer Flows - Their
  Structure and Measurement}. Oxford University Press, New York, NY, USA, 1st
  edn. (1994)

\bibitem{Kolmogorov1941b}
Kolmogorov, A.N.: {Dissipation of Energy in the Locally Isotropic Turbulence}.
  Dokl. Akad. Nauk SSSR  \textbf{32}(1) (1941). \doi{10.1038/161375b0}

\bibitem{Kolmogorov1941a}
Kolmogorov, A.N.: {The Local Structure of Turbulence in Incompressible Viscous
  Fluid for Very Large Reynolds Numbers}. Dokl. Akad. Nauk SSSR
  \textbf{30}(1),  9--13 (1941)

\bibitem{Kondrashov2016}
Kondrashov, A., Sboev, I., Dunaev, P.: {Evolution of convective plumes adjacent
  to localized heat sources of various shapes}. International Journal of Heat
  and Mass Transfer  \textbf{103},  298--304 (2016).
  \doi{10.1016/j.ijheatmasstransfer.2016.07.065}

\bibitem{Krishan2007}
Krishan, K., Kurtuldu, H., Schatz, M.F., Gameiro, M., Mischaikow, K., Madruga,
  S.: {Homology and symmetry breaking in Rayleigh-B{\'{e}}nard convection:
  Experiments and simulations}. Physics of Fluids  \textbf{19}(11), ~1--6
  (2007). \doi{10.1063/1.2800365}

\bibitem{Liu2017}
Liu, S., Shao, Y., Kunoth, A., Simmer, C.: {Impact of surface-heterogeneity on
  atmosphere and land-surface interactions}. Environmental Modelling and
  Software  \textbf{88},  35--47 (2017). \doi{10.1016/j.envsoft.2016.11.006}

\bibitem{Mellado2016}
Mellado, J.P., van Heerwaarden, C.C., Garcia, J.R.: {Near-Surface Effects of
  Free Atmosphere Stratification in Free Convection}. Boundary-Layer
  Meteorology  \textbf{159}(1),  69--95 (2016). \doi{10.1007/s10546-015-0105-x}

\bibitem{Mitzenmacher2004}
Mitzenmacher, M.: {A brief history of lognormal and power law distributions}.
  Internet Mathematics  \textbf{1}(2),  226--251 (2004)

\bibitem{Obayashi2018a}
Obayashi, I.: {Volume-Optimal Cycle: Tightest Representative Cycle of a
  Generator in Persistent Homology}. SIAM Journal on Applied Algebra and
  Geometry  \textbf{2}(4),  508--534 (2018)

\bibitem{Obukhov1941}
Obukhov, A.M.: {On the distribution of energy in the spectrum of turbulent
  flow}. Dokl. Akad. Nauk SSSR  \textbf{32}(1),  22--24 (1941)

\bibitem{Rieck2014}
Rieck, M., Hohenegger, C., van Heerwaarden, C.C.: {The Influence of Land
  Surface Heterogeneities on Cloud Size Development}. Monthly Weather Review
  \textbf{142}(10),  3830--3846 (2014). \doi{10.1175/MWR-D-13-00354.1}

\bibitem{Shao2013}
Shao, Y., Liu, S., Schween, J.H., Crewell, S.: {Large-Eddy
  Atmosphere–Land-Surface Modelling over Heterogeneous Surfaces: Model
  Development and Comparison with Measurements}. Boundary-Layer Meteorology pp.
  333--356 (2013). \doi{10.1007/s10546-013-9823-0}

\bibitem{Vereecken2016}
Vereecken, H., Pachepsky, Y., Simmer, C., Rihani, J., Kunoth, A., Korres, W.,
  Graf, A., Franssen, H.H., Thiele-Eich, I., Shao, Y.: {On the role of patterns
  in understanding the functioning of soil-vegetation-atmosphere systems}.
  Journal of Hydrology  \textbf{542},  63--86 (2016).
  \doi{https://doi.org/10.1016/j.jhydrol.2016.08.053}

\bibitem{Vuong1989}
Vuong, Q.H.: {Likelihood Ratio Tests for Model Selection and Non-Nested
  Hypotheses}. Econometrica  \textbf{57}(2),  307----333 (1989).
  \doi{10.2307/2223855}

\end{thebibliography}
\end{document}